\documentclass[twocolumn,aps,showpacs,prb,tightenlines,amsmath,amssymb,superscriptaddress]{revtex4}
\usepackage{graphicx}
\usepackage{amssymb}
\usepackage{dcolumn}
\usepackage{amsmath}
\usepackage{bm}% bold math

\usepackage{colordvi}
\newcommand{\bgreek}[1]{\mbox{\boldmath$#1$\unboldmath}}

\begin{document}

\title{Triplet-singlet relaxation in semiconductor single and 
double quantum dots}

\author{K.\ Shen}
\affiliation{Hefei National Laboratory for Physical Sciences at
Microscale,
University of Science and Technology of China, Hefei,
Anhui, 230026, China}
\author{M.\ W.\ Wu}
\thanks{Author to  whom correspondence should be addressed}
\email{mwwu@ustc.edu.cn.}
\affiliation{Hefei National Laboratory for Physical Sciences at
Microscale,
University of Science and Technology of China, Hefei,
Anhui, 230026, China}
\affiliation{Department of Physics,
University of Science and Technology of China, Hefei,
Anhui, 230026, China}
\altaffiliation{Mailing address.}

\date{\today}

\begin{abstract}
We study the triplet-singlet relaxation in two-electron semiconductor
quantum dots. Both single dots and vertically coupled double dots are
discussed. In our work, the electron-electron Coulomb interaction,
which plays an important role in the electronic structure, is
included. The spin mixing is caused by spin-orbit coupling which is
the key to the triplet-singlet relaxation. 
We show that the selection rule widely used in the literature 
is incorrect unless near the crossing/anticrossing 
point in single quantum dots. 
The triplet/singlet relaxation in double quantum dots can be markedly changed
by varying  barrier height, inter-dot distance, external
magnetic field and dot size.
\end{abstract}

\pacs{73.21.La, 71.70.Ej, 72.10.Di, 73.22.Lp}

\maketitle

\section{Introduction}
The application of semiconductor quantum dots (QDs) in generating spin-based
qubits\cite{loss1,taru} is one of the focuses in the field of
spintronics.\cite{spintronics} There are two types of qubits
investigated extensively recently.\cite{taru} One is based on the transition between
single-electron Zeeman sublevels\cite{hanson,amasha} and the other
is based on two-electron  triplet-singlet
(TS) states.\cite{petta,petta2,koppen,koppen2,sasaki,meunier}
Among these works, the decoherence time of the spin states,
including both the spin dephasing time\cite{petta,koppen} and spin
relaxation time,\cite{amasha,sasaki,meunier,petta2} has attracted much
attention as a thorough understanding of it is one of the prerequisites of the
application. There are many works on spin
relaxation reported, especially in single-electron
QDs.\cite{Woods,Sousa,wu1,dani1,Golovach, Destefani,Jose,Falko2,wu2,peter1,peter2,Westfahl}
 Recently, the TS
relaxation time of two-electron system
has also been investigated.\cite{climente,climente2,marian,dani2} It was
proposed that various mechanisms, such as the electron-phonon
scattering together with the spin-orbit coupling,\cite{dress,rashba}
the hyperfine interaction,\cite{paget,pikus}
and the cotunneling effect, could induce TS
relaxation.\cite{climente} However, the mechanism involving
electron-phonon scattering
is usually treated as the key one because the nuclei-mediated
relaxation\cite{erlingsson} and the
cotunneling can be weakened via tuning external magnetic field and tunneling
rates,\cite{sasaki} respectively.\cite{climente}
Specifically, Climente {\em et al.} used exact diagonalization technique to
calculate the two-electron spectrum structure and the phonon induced TS
relaxation in parabolic single QDs.\cite{climente} They
demonstrated the crucial role of the
excited states on spectrum structure and showed a slow decrease
of the relaxation time away from the TS
crossing in contrast to a sharp increase in the vicinity of the crossing point,
when the magnetic field is  increased from zero Tesla. This feature agrees
qualitatively  with
the recent measurement.\cite{meunier} Furthermore, their
results indicated that the spin-down 
triplet state coupled with the singlet ground state
through the spin-orbit coupling has
a much shorter lifetime compared to the other two triplet states.
This was understood by the
so called ``selection rule'' based on the perturbation
 using the lowest two single electron levels.
 Similar perturbative discussion was also given in Ref.\ \onlinecite{dani2}.
 Meunier {\em et al.} obtained perturbative wave functions from
the selection rule and treated the spin-orbit coupling coefficient as a
fitting parameter.\cite{meunier} Using these functions, they fit their
experiment data with electron-phonon-scattering-induced TS relaxation
 and obtained a particularly small spin-orbit coupling coefficient. They
attributed the reduction of the coupling
coefficient to the neglect of high excited states.
Sasaki {\em et al.} pointed out that the selection rule was correct only in the vicinity of
the TS crossing point,\cite{sasaki} which
seems to be more correct intuitively. According to the previous
work by one of the authors\cite{wu1} and confirmed by
Destefani and  Ulloa,\cite{Destefani} the spin-orbit 
coupling in quantum dots is
very strong and a large number of basis functions are
 needed in order to achieve convergence even for the lowest few states.
Therefore, whether the selection rule based on the lowest few 
levels remains unchanged when many
upper levels are involved remains questionable to us. Therefore, in
 this work we will
first reinvestigate the selection rule based on exact 
diagonalization method, jointly
with perturbation method with many basis functions.

The investigation on TS relaxation in double QD
architectures is very limited.
Recently, Wang and Wu studied the single-electron spin relaxation 
in vertically coupled double QDs and
showed that the spin relaxation can be efficiently manipulated
electronically by the inter-dot barrier.\cite{wu2} 
This suggests that the two-electron
TS relaxation should also be manipulated
by tuning inter-dot barrier height. This is another issue we
are going to explore in this work.

We organize the paper as following. In Sec.\ \ref{model} we set up the
 model and lay out the formalism.
 Then in Sec.\ \ref{result} we show our numerical results. We discuss the
single dot case in Sec.\ \ref{singledot}. We first show the exact
diagonalization results with sufficient basis functions.
We then reexamine the selection rule by using more basis functions
instead of the lowest two, both perturbatively and exactly.
We show the selection rule widely used in the literature is not correct
except near the TS crossing/anticrossing points.
In Sec.\ \ref{doubledot}, we show the results of double QDs.
We summarize in Sec.\ \ref{sum}.

\section{Model and formalism}
\label{model}
We start our investigation from a 
vertically-coupled double QD:   Electrons are confined
by a parabolic potential
$V_c(x,y)=\frac{1}{2}m^{\ast}\omega_0^2(x^2+y^2)$ (corresponding to the
effective dot diameter $d_0=\sqrt{\hbar\pi/m^{\ast}\omega_0}$) along
the $x$-$y$ plane,\cite{fock,darwin} with $m^{\ast}$ representing the effective
mass. Along the $z$-axis, a strong confinement is given by
\begin{equation}
  V_z(z)=
\begin{cases}
  V_0\ , & \qquad
  |z|\leqslant \frac{1}{2}a\ , \\
   0\ , & \qquad
   \frac{1}{2}a<|z|<\frac{1}{2}a+d\ ,\\ \infty\ , & \qquad
   \mbox{otherwise}\ ,
\end{cases}
\label{eq1}
\end{equation}
with $V_0$, the inter-dot barrier.\cite{austing}
By taking $a=0$, one comes to the single dot configuration.
The single-electron Hamiltonian with magnetic field along the growth direction ($z$) is given by
\begin{equation}
  H_e=\frac{\bf P^{2}}{2m^{\ast}}+V({\bf r})+H_{so}({\bf P})+H_Z\ ,
\label{eq2}
\end{equation}
in which $V({\bf r})=V_z(z)+V_c(x,y)$ and
${\bf P}=-i\hbar{\mbox{\boldmath $\nabla$\unboldmath}}+e/c{\bf A}$ with ${\bf
A}=(B/2)(-y,x,0)$. $H_{so}$ represents the spin-orbit coupling which
is the key to the spin flip. In this work, we only consider the Dresselhaus spin-orbit
coupling\cite{dress} as the Rashba coupling\cite{rashba} is comparably small in GaAs QDs.\cite{lau}
Hence $H_{so}=\gamma\bf h\cdot\bgreek\sigma$, with
${\bf h}=[P_x(P_y^2-P_z^2),P_y(P_z^2-P_x^2),P_z(P_x^2-P_y^2)]$.\cite{yakonov}
For small well width, it reduces to
\begin{equation}
  H_{so}=\frac {\gamma}{\hbar^3}
  \langle P_{z}^{2}\rangle(-P_x\sigma_x+P_y\sigma_y)\ ,
\label{eq3}
\end{equation}
with $\langle P_z^2\rangle$ the average of $P_z^2$ over the electronic
states defined by $V_z(z)$.
$H_Z=\frac{1}{2}g\mu_{B}B\sigma_z$ is the Zeeman splitting with
$g$ being the Land\'{e} factor. We define
$H_0=\frac{\bf P^{2}}{2m^{\ast}}+V({\bf r})$, whose
eigenvalues and eigenfunctions can be obtained from
the Schr\"{o}dinger equation
\begin{equation}
  H_0|\phi\rangle=E|\phi\rangle\ .
\label{eq4}
\end{equation}
Previous work on single-electron QDs gives the solution of the lateral part of
this equation,\cite{fock,darwin,wu1} where the exact energy levels are given by
\begin{equation}
  E_{nl}=\hbar\Omega(2n+|l|+1)+\hbar l\omega_B\ ,
\label{eq5}
\end{equation}
with $\Omega=\sqrt{\omega_{0}^2+\omega_{B}^2}$ and
$\omega_B=eB/(2m^{\ast})$. The wave functions read
\begin{equation}
  \langle {\bf r}|nl\rangle=N_{n,l}(\alpha r)^{|l|}e^{-(\alpha
    r)^2/2}L_{n}^{|l|}((\alpha r)^2)e^{il\theta}\ ,
\label{eq6}
\end{equation}
with $N_{n,l}=(\alpha^{2}n!/\pi (n+|l|)!)^{1/2}$ and
$\alpha =\sqrt{m^{\ast}\Omega/\hbar}$. $L_{n}^{|l|}$ is the generalized
Laguerre polynomial. In these equations,
$n=0,1,2,...$ is the radial quantum number and $l=0,\pm 1,\pm 2,...$
is the azimuthal angular momentum quantum number. By solving the
$z$-component of Eq.\ (\ref{eq4}), we obtain the lowest two  electronic states
along the $z$-axis as following:
\begin{equation}
  \phi_z^0=
  \begin{cases}
    C_1^0\sin[k(z-\frac{a}{2}-d)]\ , & \frac{a}{2}<z<\frac{a}{2}+d\ ,\\
    C_2^{0}\cosh(\beta z)\ , &  |z|\leqslant \frac{1}{2}a\ , \\
    C_1^0\sin[k(-z-\frac{a}{2}-d)]\ , & -\frac{a}{2}-d<z<-\frac{a}{2}\ ,
  \end{cases}
\label{eq7}
\end{equation}
and
\begin{equation}
  \phi_z^1=
  \begin{cases}
    C_1^1\sin[k(z-\frac{a}{2}-d)]\ , & \frac{a}{2}<z<\frac{a}{2}+d\ ,\\
    C_2^1\sinh(\beta z)\ , & |z|\leqslant \frac{1}{2}a\ ,\\
    C_1^1\sin[k(z+\frac{a}{2}+d)]\ , & -\frac{a}{2}-d<z<-\frac{a}{2}\ ,
  \end{cases}
\label{eq8}
\end{equation}
in which $k^2={2m^\ast E_z}/{\hbar^2}$ and $\beta^2={2m^\ast
  (V_0-E_z)}/{\hbar^2}$ with $E_z$ denoting the energy along
this direction.
We use the superscripts ``0'' and ``1'' to denote the even and odd parity respectively.
The total spatial  wave function is then denoted by
$|nln_z\rangle$, with $n_z=0$ and 1 in this work to distinguish the above
even and odd states along the $z$-axis. Due to the strong confinement along the
$z$-axis, levels higher than $n_z=1$ are neglected.
It is noted that  when we refer to the single QDs, we only keep the lowest state (the even one)
due to the small well width.

For two-electron system, the total Hamiltonian is
written as
\begin{equation}
   H_{tot}=(H_e^1+H_e^2+H_{C})+H_{ep}^1+H_{ep}^2+H_p\ .
\label{eq9}
\end{equation}
In this equation, the third term $H_{C}=\frac{e^2}{4\pi\epsilon_0\kappa \bf|r_1-r_2|}$
describes the Coulomb interaction between the two electrons with $\kappa$
representing the static dielectric constant. $H_p=\Sigma_{{\bf q
    }\lambda}\hbar\omega_{{\bf q} \lambda}a_{{\bf q}
  \lambda}^{+}a_{{\bf q} \lambda}$ represents the phonon Hamiltonian, and
$H_{ep}=\Sigma_{{\bf q} \lambda}M_{{\bf q} \lambda}(a_{{\bf q}
  \lambda}^{+}+a_{{\bf q} \lambda})\exp(i\bf q \cdot r) $ is the Hamiltonian of
the electron-phonon interaction.
The superscripts ``1'' and ``2'' label the two electrons.

We construct two-electron basis functions from 
the single electron wave functions.
To see the physics clearly, we
construct our two-electron basis functions in either singlet or
triplet forms.
Taking two single-electron spatial wave functions $|n_1l_1n_{z1}\rangle$ and
  $|n_2 l_2 n_{z2}\rangle$ (denoted as $|N_1\rangle$ and
$|N_2\rangle$ for short) as an
example, the singlet functions can be constructed by
\begin{eqnarray}
  \nonumber
  |S\rangle&=&(|\uparrow\downarrow\rangle-|\downarrow\uparrow\rangle)\\
  &&\otimes
  \begin{cases}
    \frac{1}{\sqrt 2}|N_1N_2\rangle\ ,& N_1=N_2\ ,\\
    \frac{1}{2}(|N_1N_2\rangle+|N_2N_1\rangle)
    \ ,& N_1\not=N_2\ ,
  \end{cases}
\label{eq10}
\end{eqnarray}
and the triplet functions for $N_1\neq N_2$ by
\begin{eqnarray}
  \label{eq11}
  &&|T_+\rangle=\frac{1}{\sqrt2}(|N_1N_2\rangle-|N_2N_1\rangle)\otimes|\uparrow\uparrow\rangle\ ,\\
  &&|T_0\rangle=\frac{1}{2}(|N_1N_2\rangle-|N_2N_1\rangle)\otimes(|\uparrow\downarrow\rangle
  +|\downarrow\uparrow\rangle)\ ,\\
  &&|T_-\rangle=\frac{1}{\sqrt2}(|N_1N_2\rangle-|N_2N_1\rangle)\otimes|\downarrow\downarrow\rangle\ .
  \label{eq13}
\end{eqnarray}
Here, $N$ and $N^\prime$, in the ket $|NN^\prime\rangle$, represent
the spatial  quantum numbers of the first and
the second electrons respectively. We define the total angular
  momentum $L=l_1+l_2$ and denote the total spin $(S,S_z)$ with $S_z$
  representing the $z$-component of the total spin ${\bf S}$.

Then, we calculate the matrix elements of the Coulomb interaction and the spin-orbit coupling\cite{wu1}
under these basis functions. 
The Coulomb matrix elements can be expressed in the  form
\begin{eqnarray}
  \nonumber
\langle N_1N_2|H_C|N_1^\prime N_2^\prime\rangle&=&
\frac{e^2}{4\pi^2\epsilon_0\kappa}\delta_{l_{N_1}+l_{N_2},l_{N_1^\prime}+l_{N_2^\prime}}\\
  &&\mbox{}\times Q(N_1,N_2,N_1^\prime,N_2^\prime)\ ,
\label{eq14}
\end{eqnarray}
in which $Q$ is given in detail in Appendix\ \ref{appen}.
Thus we obtain the two-electron Hamiltonian.
By diagonalizing the two-electron Hamiltonian, 
one obtains all the energy levels and
eigenfuctions. We identify a state as singlet/triplet  if its
amplitude of singlet/triplet components is larger than 50\ \%.
We rewrite the  spin-orbit coupling Hamiltonian
[Eq.\ (\ref{eq3})] using the ladder operators as\cite{climente}
\begin{equation}
  H_{so}=\gamma_c(P^+S^++P^-S^-)\ ,
\label{eq15}
\end{equation}
with the coupling coefficient $\gamma_c=\frac {\gamma}{\hbar^3}
  \langle P_{z}^{2}\rangle$.
Then it is noted that $P^{\pm}$ and $S^{\pm}$ change
$L$ and $S_z$ by one unit, respectively.
It suggests that a state with $(L, S_z)$
can only be coupled with the states with $(L+1, S_z+1)$ and $(L-1,
S_z-1)$. 

Treating $|i\rangle$ and $|f\rangle$ as the
initial and final states, we can calculate the phonon-induced
 relaxation rate from the Fermi Golden Rule
\begin{eqnarray}
  \nonumber
  \Gamma_{i\rightarrow f}&=&\frac{2\pi}{\hbar}\sum_{{\bf q}\lambda}|
  {M}_{{\bf q}\lambda}|^2|\langle f|\chi|i\rangle|^2[\bar{n}_{{\bf
      q}\lambda}\delta(\epsilon_f-\epsilon_i-\hbar\omega_{{\bf q}\lambda}) \\
  && +(\bar{n}_{{\bf
      q}\lambda}+1)\delta(\epsilon_f-\epsilon_i+\hbar\omega_{{\bf
      q}\lambda})]\ ,
\label{eq16}
\end{eqnarray}
in which $\chi({\bf q,r_1,r_2})=e^{i{\bf
  q} \cdot {\bf r}_1}+e^{i{\bf q} \cdot {\bf r}_2}$ comes from the total
electron-phonon interaction Hamiltonian
$H_{ep}=H_{ep}^1+H_{ep}^2$. Here, $\bar{n}_{{\bf q}\lambda}$
represents the Bose distribution of
phonon with mode $\lambda$ and momentum $\bf q$. In our calculation,
the temperature is fixed at 0\ K. Therefore only the phonon emission process occurs.

\section{Numerical results}
\label{result}
In the numerical calculation, we include the electron-acoustic
 phonon scattering
due to the deformation potential with $|{M}_{{\bf q}s l}|^2=\hbar\Xi
^2q/2Dv_{sl}$,\cite{vogl}and  due
to the piezoelectric field with $|{M}_{{\bf q}pl}|^2
=(32\hbar\pi^2e^2e_{14}^{2}/\kappa^2Dv_{sl})[(3q_xq_yq_z)^2/q^7]$
for the longitudinal mode\cite{mahan} and
$\Sigma_{j=1,2}|M_{{\bf
    q}pt_j}|^2=(32\hbar\pi^2e^2e_{14}^2/\kappa^2Dv_{st}q^5)[q_{x}^{2}q_{y}^{2}
+q_{y}^{2}q_{z}^{2}+q_{z}^{2}q_{x}^{2}-(3q_xq_yq_z)^2/q^2]$
for the two transverse modes.\cite{lei} Here, $\Xi=7$\ eV stands for the acoustic
deformation potential; $D=5.3\times 10^3$\ kg/m$^3$ is the GaAs volume density;
$e_{14}=1.41\times 10^9$\ V/m denotes the piezoelectric constant and the static dielectric
constant $\kappa$ is 12.9; $v_{sl}=5.29\times10^3$\ m/s corresponds to the longitudinal sound velocity
and $v_{st}=2.48\times 10^3$\ m/s corresponds to the transverse one.\cite{bs}

In our calculation,  $g$ factor is
$-0.44$\cite{bs,hanson} and the Dresselhaus
coefficient $\gamma$ is 21.5\ \AA$^3\cdot$eV.\cite{richards}
The typical electron effective mass $m^{\ast}$ in GaAs is $0.067m_0$,\cite{bs}
with $m_0$ being the free electron mass.
\begin{figure}[bth]
\centering
%\begin{center}
\includegraphics[height=5.cm]{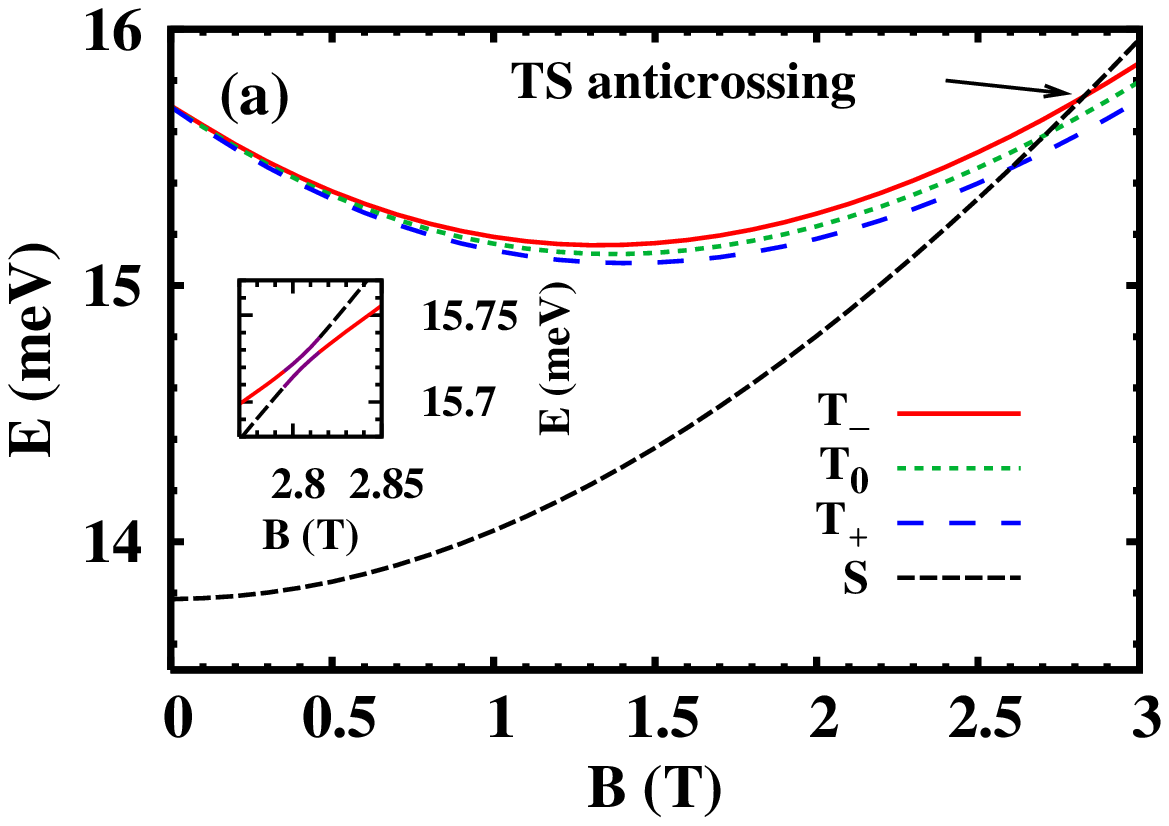}
\includegraphics[height=5.cm]{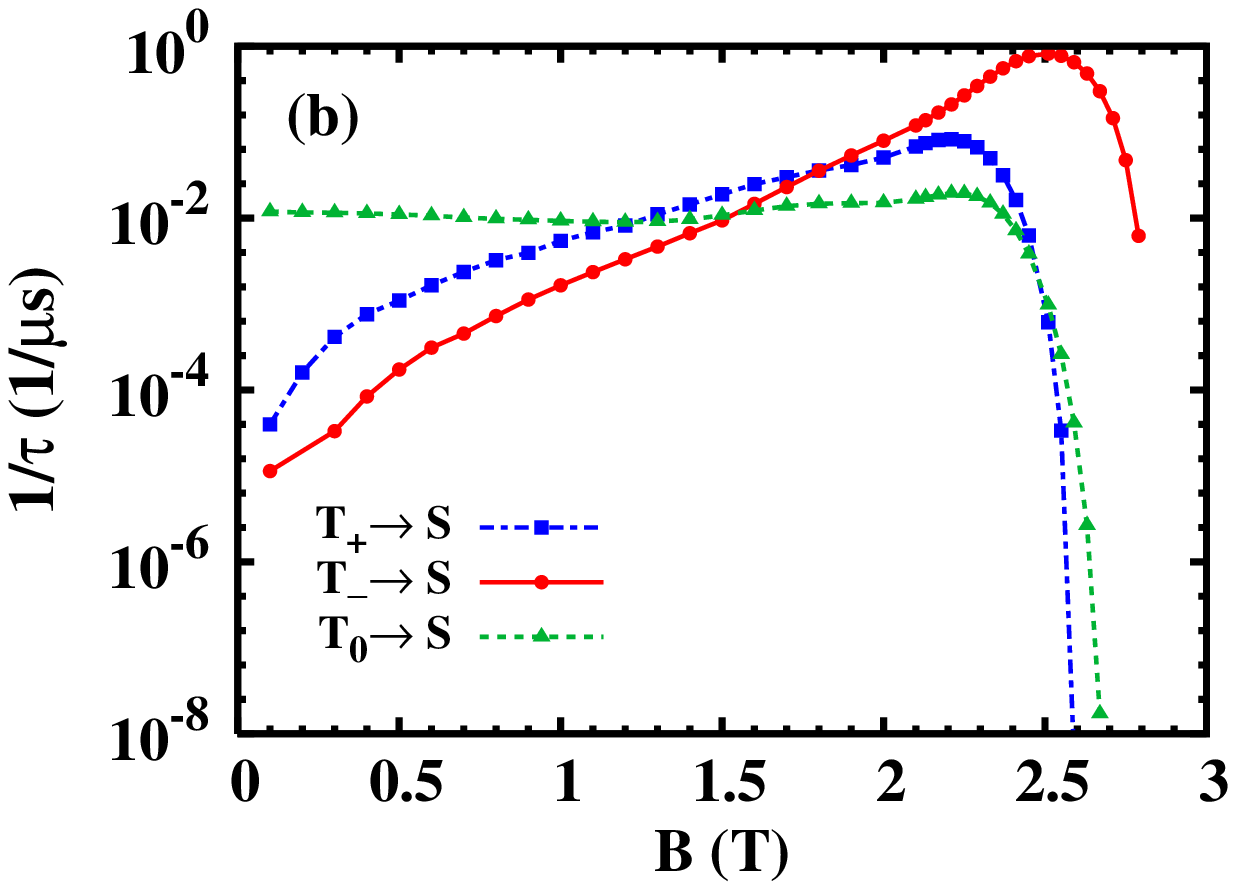}
%\end{center}
\caption{(Color online) (a) The lowest four energy levels {\em vs.} magnetic
  field $B$ in single QD. The TS anticrossing point between 
 $T_-$ and $S$ is shown and the range near this point is enlarged
  in the inset. (b) $\tau^{-1}$ of the three transition channels {\em
  vs.} the magnetic field. In the calculation,
 $d=$5\ nm and $d_0=30$\ nm.}
\label{fig1}
\end{figure}

\subsection{Single dot}
\label{singledot}

We first set $a=0$ to  investigate the single dot case
by exact diagonalization method with the lowest 800 singlet and 2220
triplet basis functions. Under the basis, the energy levels and the 
TS relaxation rates
  are well converged. The magnetic field
dependence of the first four levels and that of the
TS relaxation rates are plotted in Fig.\ \ref{fig1}. In the
calculation, we take the well width\ $2d=10$\ nm and the
effective diameter $d_0=30$\ nm. From Fig.\ \ref{fig1}(a),
one notices that the ground state is a singlet denoted as ``S'', 
in a wide range of the
magnetic field (from $0$\ T to $2.6$\ T approximately). In this region, the first
three excited states are triplet states, labeled as  $|T_+\rangle$
(spin-up), $|T_0\rangle$ (spin-zero) and $|T_-\rangle$ (spin-down), 
and the energy of
$|T_-\rangle$ is the highest one among the three because of the
Zeeman effect. When the magnetic field increases from $2.6$\ T, one further observes a
TS  crossing  between the
singlet and the two triplets ($|T_+\rangle$ and $|T_0\rangle$). Moreover, a 
TS anticrossing point
(with a small  energy gap shown in
 Fig.\ {\ref{fig1}(a)}) 
also exists between the singlet and $|T_-\rangle$ triplet state due to the
Dresselhaus spin-orbit coupling.  
From the calculation, we notice that the major components of
  $|S\rangle$, $|T_\pm\rangle$ and $|T_0\rangle$ are $|S^1\rangle$,
$|T^1_\pm\rangle$ and $|T^1_0\rangle$, which are the lowest
  singlet and triplet basis functions. Specifically, using
 the lowest two single-particle wave functions, 
 $|nln_z\rangle$ with $n=n_z=0$, $l=0$ and $-1$, one can construct
 $|S^1\rangle$ with  
$|000\rangle$ and $|000\rangle$, and $|T^1_\pm\rangle$ and $|T^1_0\rangle$
 with $|000\rangle$ and $|0-10\rangle$ according to Eqs.\
(\ref{eq10})-(\ref{eq13}). 
Therefore, the quantum numbers $(L,S_z)$ of $|S^1\rangle$, $|T^1_+\rangle$, 
$|T^1_0\rangle$ and 
$|T^1_-\rangle$ are different, i.e.,
$(0,0)$, $(-1,1)$, $(-1,0)$ and $(-1,-1)$ respectively.

From Fig.\ \ref{fig1}(b) one observes that the TS relaxation rates
increase slowly with the magnetic field away from the 
crossing/anticrossing points,
but decrease dramatically  in the vicinity of the crossing/anticrossing
points, in agreement with the measurement
qualitatively.\cite{meunier} The relaxation rate reaches maximum
where the wave length of the emissive phonon is
comparable with the dot size.\cite{scat} In our calculation, the
TS splitting, i.e., the energy  between
the triplet and the singlet, $\Delta_{TS}\sim0.2$\ meV.
The corresponding half-wavelength of the 
transverse phonon is therefore 
 about 30\ nm as the dot diameter $d_0$. This feature was interpreted as the
competing  effects of the magnetic field on the electron-phonon
coupling and the spin-orbit coupling.\cite{climente} Actually, the
strength of the spin-orbit coupling is proportional to $\alpha$ [see Eq.\ (\ref{eq6})]
which increases with the magnetic field,\cite{wu1} whereas the
electron-phonon scattering becomes rather weak when the emissive
phonon momentum decreases.\cite{climente}

Surprisingly, our results are very different from
those shown in the previous work,
 where the transition rate of
$|T_-\rangle$ is much larger than those of the other two triplet
states $|T_+\rangle$ and $|T_0\rangle$.\cite{climente} In that work,
the authors interpreted their
results by the selection rule based on the perturbation method
including the lowest four basis functions, i.e., $|S^1\rangle$,
$|T^1_+\rangle$, $|T^1_0\rangle$ and $|T^1_-\rangle$. 
Under that basis, only $|T_-\rangle$ is coupled with $|S\rangle$ 
through the Dresselhaus
spin-orbit coupling according to 
Eq.\ (\ref{eq15}). So only the transition from $|T_-\rangle$ to $|S\rangle$ can
occur. Thus they concluded that the transition
rate from $|T_-\rangle$ to $|S\rangle$ is much
larger than those of the other channels even though much more 
 (instead of four) basis functions
are included.  In fact, this selection 
rule is widely used in the literature.\cite{sasaki,meunier}
However, as one needs many basis
functions to achieve convergence even in the single-electron QD
system,\cite{wu1} whether the selection rule from the lowest four 
basis functions
is robust against the inclusion of higher basis functions
 remains an open question.
Here we reexamine the selection rule with more
basis functions. Assuming the perturbation based on
the lowest four states $|S^1\rangle$, $|T_\pm^1\rangle$ and $|T_0^1\rangle$
 is adequate to
describe the real physics, we expect the selection rule should always
be valid when more basis functions are included.
Specifically, we now use four single-electron functions $|000\rangle$,
$|0-10\rangle$, $|010\rangle$ and $|0-20\rangle$ to construct the
two-electron basis functions. Keeping only the index of $l$ 
from $|nln_z\rangle$ since the other two are fixed, the
six lowest singlet states are constructed by 
$|0\rangle|0\rangle$, $|0\rangle|-1\rangle$,
$|0\rangle|1\rangle$, $|-1\rangle|-1\rangle$ , $|0\rangle|-2\rangle$, 
and $|-1\rangle|1\rangle$ separately and the three lowest triplet states
 are constructed by $|0\rangle|-1\rangle$ 
in the way of Eqs.\ (\ref{eq10})-(\ref{eq13}). We
denote these nine basis functions as
$|S^1\rangle$, $|S^2\rangle$, $|S^3\rangle$, $|S^4\rangle$,
$|S^5\rangle$, $|S^6\rangle$, $|T^1_+\rangle$,
$|T^1_0\rangle$ and $|T^1_-\rangle$ in sequence, and the quantum
 numbers $(L,S_z)$ are 
  $(0,0)$, $(-1,0)$, $(1,0)$, $(-2,0)$, $(-2,0)$, $(0,0)$, $(-1,1)$,
  $(-1,0)$, and $(-1,-1)$ respectively. Therefore, only the singlet states
$|S^1\rangle$ and $|S^6\rangle$ can mix with $|T^1_-\rangle$;
$|S^4\rangle$ and $|S^5\rangle$ can mix with $|T^1_+\rangle$, according to
Eq.\ (\ref{eq15}) under these basis functions. 
No mixing occurs to the state $|T^1_0\rangle$.

\begin{figure}[bth]
\centering
\includegraphics[height=5.8cm]{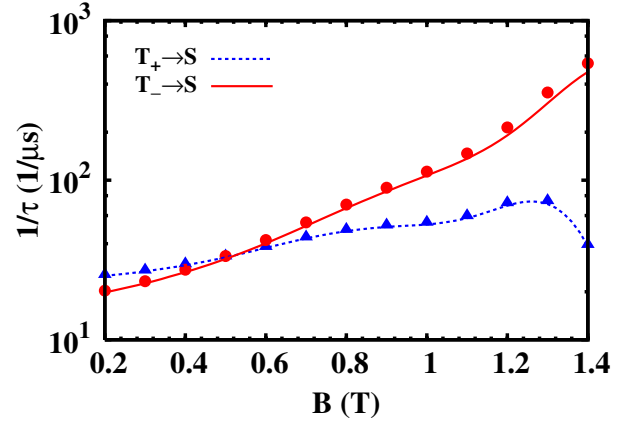}
\caption{(Color online) Comparison of the TS relaxation rates
of $T_-$ and $T_+$ in single QD calculated from the perturbation method and the
exact diagonalization method limited 
within  nine basis functions ($|S^1\rangle$ to $|S^6\rangle$,
$|T_\pm^1\rangle$ and $|T_0^1\rangle$). 
Dashed curve and  $\blacktriangle$: $1/\tau_{T_+\rightarrow S}$ from the
exact diagonalization and perturbation methods respectively; 
Solid curve and $\bullet$: $1/\tau_{T_-\rightarrow S}$ from the
exact diagonalization and perturbation methods.
}
\label{fig2}
\end{figure}

As the Coulomb interaction is too strong
to treat perturbatively, we first diagonalize the
Hamiltonian with the Coulomb interaction included 
to obtain a new set of basis functions, i.e.,
$|\bar S^1\rangle=a_1|S^1\rangle+b_1|S^6\rangle$,
$|\bar S^2\rangle=a_2|S^1\rangle+b_2|S^6\rangle$,
$|\bar S^3\rangle=a_3|S^4\rangle+b_3|S^5\rangle$,
$|\bar S^4\rangle=a_4|S^4\rangle+b_4|S^5\rangle$,
$|\bar S^5\rangle=|S^2\rangle$, $|\bar S^6\rangle=|S^3\rangle$,
$|\bar T_+^1\rangle=|T^1_+\rangle$,
$|\bar T_0^1\rangle=|T^1_0\rangle$, and
$|\bar T_-^1\rangle=|T^1_-\rangle$. Here $a_i$ and $b_i$ are obtained
from the numerical diagonalization. The corresponding
eigenvalues are $E_1$ to $E_6$, $E_+$,
 $E_0$ and $E_-$, respectively. Then we treat the spin-orbit coupling
as perturbation under the new basis functions. The
  lowest four states then read
\begin{eqnarray}
  &&|T_+\rangle=|T^1_+\rangle+\theta_+^1|S^4\rangle+\theta_+^2|S^5\rangle\
  ,\label{eq17}\\
  &&|T_0\rangle=|T^1_0\rangle\ ,\\
  &&|T_-\rangle=|T^1_-\rangle+\theta_-^1|S^1\rangle+\theta_-^2|S^6\rangle\ ,\label{eq19}\\
  &&|S\rangle=\theta_s^1|S^1\rangle+\theta_s^2|S^6\rangle+\theta_s^3|T^1_-\rangle\ ,
\label{eq20}
\end{eqnarray}
with $\theta_+^{1(2)}=\sum_{i=3,4}\frac{(b_i^{\ast}-a_i^{\ast})
      {\mathcal A}}{E_+-E_i} {\Theta^{1(2)}_i}$, $\theta_s^{1(2)}=\Theta^{1(2)}_1$,
    $\theta_-^{1(2)}=\sum_{i=1,2}\frac{a_i^{\ast}{\mathcal 
      A}+b_i^{\ast}{\mathcal B}}{E_--E_i}\Theta^{1(2)}_i$
  and $\theta_s^3=\frac{a_1{\mathcal A}+b_1{\mathcal B}}{E_--E_1}$.
Here $\Theta^{1(2)}_i=a_i(b_i)$, ${\mathcal A}=-i\gamma^{\ast}\alpha(1-eB/2\hbar\alpha^2)$
  and ${\mathcal B}=-\frac{i}{\sqrt
    2}\gamma^{\ast}\alpha(1+eB/2\hbar\alpha^2)$ with $\gamma^\ast$
 being $\gamma(\pi/2d)^2$. 

Obviously, the transitions from both $|T_+\rangle$ and $|T_-\rangle$ to $|S\rangle$ can occur
according to Eqs.\ (\ref{eq17})-(\ref{eq20}). 
The matrix elements $|\langle f|\chi|i\rangle|^2$ in Eq.\
(\ref{eq16}) now read
\begin{eqnarray}
  \nonumber
  \label {eq21}
  &&|\langle S|\chi|T_+\rangle|^2=|{\theta_s^1}^\ast{\theta_+^2}\langle
  S^1|\chi|S^5\rangle +{\theta_s^2}^\ast{\theta_+^1}\langle
  S^6|\chi|S^4\rangle|^2\\
  &&\hspace{1.83cm}=(xt)^2|\xi|^2 I^2(q_z)\ ,\\
  \nonumber
  &&|\langle S|\chi|T_-\rangle|^2=|{\theta_s^1}^\ast{\theta_-^1}\langle
  S^1|\chi|S^1\rangle +{\theta_s^2}^\ast{\theta_-^2}\langle
  S^6|\chi|S^6\rangle\\
  \nonumber
  &&\hspace{2.2cm}+{\theta_s^3}^\ast\langle T^1_-|\chi|T^1_-\rangle|^2 I^2(q_z)\\
  &&\hspace{1.83cm}=|2t\zeta_1-tx\zeta_2|^2 I^2(q_z)\ ,
  \label {eq22}
\end{eqnarray}
with $x=k_{\|}^2/4\alpha^2$, $t=e^{-x}$, 
$I(q_z)=\pi^2\sin(dq_z)/\{dq_z[\pi^2-(dq_z)^2]\}$,
$\zeta_1={\theta_s^3}^\ast+{\theta_s^1}^\ast\theta_-^1
+{\theta_s^2}^\ast\theta_-^2$, $\zeta_2={\theta_s^3}^\ast+2
{\theta_s^2}^\ast\theta_-^2$ and $\xi={\theta_s^1}^\ast
\theta_+^2+\sqrt 2{\theta_s^2}^\ast\theta_+^1$. We calculate the
relaxation rates of these two channels and plot the results
in Fig.\ \ref{fig2}. One notices that the
two sets of dots (blue $\blacktriangle$ for
$|T_+\rangle$ and red $\bullet$ for
$|T_-\rangle$)  are quite close to each other and
even show a crossing. In other words, the selection rule
 is violated. We also present the exact
diagonalization results under the same basis functions 
$|S^1\rangle$-$|S^6\rangle$, 
$|T^1_0\rangle$ and $|T^1_\pm\rangle$ in
Fig.\ \ref{fig2} (blue dashed curve for
$|T_+\rangle$ and red solid curve for
$|T_-\rangle$). It is seen that the diagonalization results almost
exactly match the perturbation results. 
This match further confirms that both our exact
diagonalization and the perturbation calculations are correct.
Compare Fig.\ \ref{fig2} with Fig.\ \ref{fig1}(b), it is obvious that the
high excited levels manifest themselves markedly in the relaxation rates.
From our calculation, we notice that the coefficients in Eqs.\ (\ref{eq17}) and
(\ref{eq19}) are comparable. This is because that the 
denominators $E_--E_i$ in
$\theta_-^{1(2)}$ are close to $E_+-E_i$ in
$\theta_+^{1(2)}$.
This explains the reason why the curve of $|T_-\rangle$ is close
to that of $|T_+\rangle$ in Fig.\ \ref{fig2}.

However, it is noted that the selection rule works well in the
vicinity of the crossing/anticrossing points both in Figs.\
\ref{fig1}(b) and \ref{fig2}.\cite{efig2} This can be understood from Eqs.\
(\ref{eq17}) and (\ref{eq19}).  Near the TS crossing point where
$E_+\sim E_1$, the energy splitting $E_+-E_{3,4}$ is finite. Therefore
$\theta_+^{1(2)}$ only changes slightly compared with the region away
from the TS crossing.  Similar is true for the coefficients of
$|T_0\rangle$.  In contrast, $\theta_-^{1(2)}$ is very large when
$E_-\sim E_1$. Therefore the transition rate from $|T_-\rangle$ would
be much larger than those from $|T_+\rangle$ and $|T_0\rangle$, i.e.,
the selection rule is valid in the vicinity of the TS
crossing/anticrossing point. Moreover, the effect of the Zeeman
splitting also makes the transition rate of $|T_-\rangle$ larger than
those of $|T_+\rangle$ and $|T_0\rangle$ because of the larger phonon
momentum $q$. Specifically, the energy splitting between $|T_-\rangle$
and $|S\rangle$ is about $0.18$\ meV at $B=2.5$\ T in Fig.\ \ref{fig1}(a),
which is much larger than that between $|T_+\rangle$ ($|T_0\rangle$)
and $|S\rangle$, i.e., $\sim 0.06$\ meV (0.12\ meV). As the transition rates
are proportional to $q^m$ with $m>0$ varying for different mechanisms, 
the rate of $|T_-\rangle$
is much larger than those of $|T_+\rangle$ and $|T_0\rangle$.

\subsection{Double dot}
\label{doubledot}

Now we turn to study the TS relaxation rate in weakly coupled double QDs
using a basis functions including 400 singlet and 1080 triplet states.
In the calculation, $a=8$\ nm and $d=7$\ nm. 
In this part we still use $|T_\pm\rangle$ and $|T_0\rangle$
($|S\rangle$) to denote eigenfunctions of the lowest three
triplet states (lowest singlet state). 
To determine the contribution of the energy levels
 along the $z$-axis, we take the barrier
  height $V_0=0.25$\ meV,  the lowest one in our calculation, as
an example. In this configuration the
 splitting between the first and the second levels along the $z$-axis is about
$1$\ meV  and that between the second and the third levels is much
 larger, about 0.2\ eV. Compared with the lateral confinement ($\sim
  4$\ meV for $d_0=30$\ nm), we only need to include the lowest two in our
calculation. 

\begin{figure}[bth]
\begin{center}
\includegraphics[height=6cm]{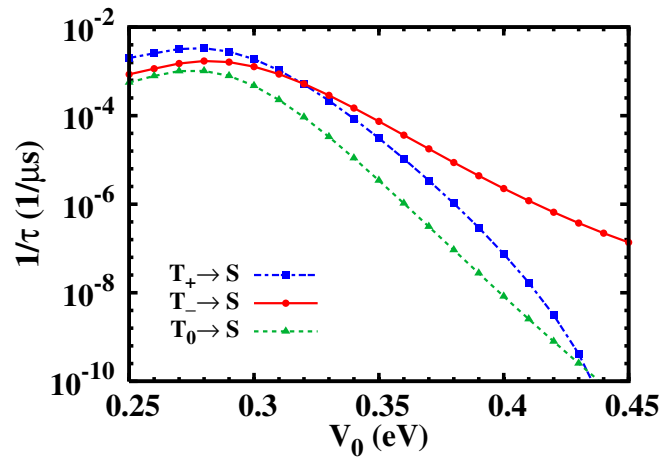}
\end{center}
\caption{(Color online) $\tau^{-1}$  {\em vs.} barrier height in double
  QDs. In the calculation, $a=8$\ nm, $d=7$\ nm, $d_0=30$\
  nm and $B=0.5$\ T.}
\label{fig4}
\end{figure}

We first investigate the TS relaxation rate as a function
of the barrier height. In the calculation, $d_0=30$\
  nm and $B=0.5$\ T. As shown in Fig.\ \ref{fig4},
each transition rate first
increases slowly until it reaches the maximum around $V_0\sim0.28$\ eV where
the TS splitting $\Delta_{TS}\sim0.4$\ meV corresponding 
to the wavelength of the emissive
phonon being comparable with the dot size $d_0$.\cite{scat}
After that, the TS relaxation rate decreases rapidly with the barrier height.
 This would offer us a scheme to manipulate the TS
relaxation in double QDs. Similar features (not
shown here) are obtained when we increase the inter-dot distance. The
dramatic decrease of the relaxation rate can be understood as following. 
When the barrier height becomes higher
or the inter-dot distance becomes larger, the inter-dot coupling is weakened
and the energy splitting between the lowest two levels along the $z$-axis
becomes  smaller. As a result, the splitting between $|T_\pm\rangle$
($|T_0\rangle$)  and $|S\rangle$ decreases too.
This causes the decrease of the TS relaxation rate as discussed in
the previous subsection
.
\begin{figure}[bth]
\begin{center}
\includegraphics[height=6cm,width=8.5cm]{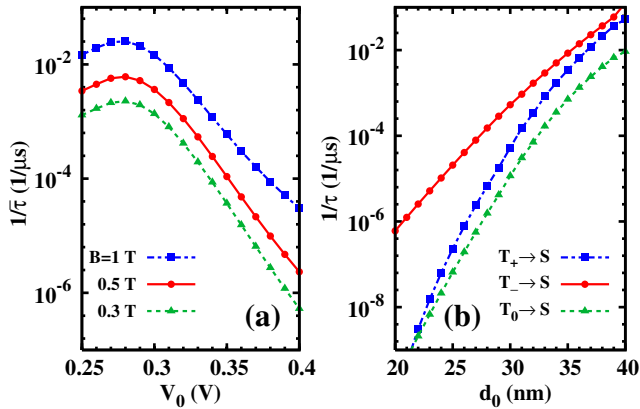}
\end{center}
\caption{(Color online) (a) Average TS relaxation rate 
$\bar\tau^{-1}$  {\em vs.} barrier
 height at different  magnetic fields in double QDs. 
In the calculation, $a=8$\ nm, $d=7$\ nm and $d_0=30$\ nm. (b) $\tau^{-1}$
  {\em vs.} effective dot diameter. In the calculation, $a=8$\ nm, $d=7$\ nm,
  $V_0=0.35$\ V and $B=1$\ T.}
\label{fig5}
\end{figure}

To have a look at the role of the
magnetic field, we calculate the average relaxation rate
 $1/\bar\tau=(1/\tau_{T_+\rightarrow S}+1/\tau_{T_-\rightarrow S}+1/\tau_{T_0\rightarrow S})/3$ as function
of the barrier height at different magnetic
field in Fig.\ \ref{fig5} (a), but with the  dot size $d_0=30$\ nm fixed. 
It is seen from the figure that  higher
magnetic field leads to relatively larger transition rate. It is due to
the enhanced spin-orbit coupling in strong magnetic field.
The influence of the effective diameter of QDs with $V_0=0.35$\ V and
$B=1$\ T is also shown in Fig.\ \ref{fig5}(b). One
finds the transition rates increase with the effective diameter
$d_0$. The reason lies on the different symmetry properties of the singlet
and triplet states. For the singlet state, the inter-electron distance
decreases with the decrease of the dot size. The coulomb repulsion therefore
lifts the corresponding energy levels. However, the energy 
lifts of the triplet states are 
 smaller due to the antisymmetry property of the triplet states
which  prevents the electrons to be close to each other. Therefore, the 
TS splitting becomes smaller with the decrease of the dot size.
This leads to the rapid decrease of the TS relaxation rates.

\section{Summary}
\label{sum}
In summary, we have investigated the TS relaxation
in single and double QDs. For the single dot case, we
find that the average relaxation rate first slowly 
increases with magnetic field until it
reaches the maximum where the wavelength of emissive phonon
is comparable with the dot size. Then it 
drops sharply. This result qualitatively
agrees with the recent
measurement.\cite{meunier} Furthermore, our result 
shows the transition rates of the
  triplet $|T_+\rangle$ and $|T_0\rangle$ can be comparable with that of
 $|T_-\rangle$, which violates the selection rule
in the literature.\cite{climente} We show that the 
selection rule obtained from the lowest four basis functions does
not hold in general cases where much more basis functions are needed to
converge the triplet/singlet states. This is shown perturbatively by
calculating the TS relaxation rates based on nine basis functions. 
Comparable  transition rates of
 $|T_+\rangle$ and $|T_-\rangle$ are immediately obtained 
away from the TS crossing point. 
The perturbation results are in good agreement with the
exact diagonalization results under the same
 basis functions.  We also show
 that the selection rule works well in the vicinity of the TS
 crossing/anticrossing point due to the effects from the
 Zeeman splitting and the anticrossing.
 For the double QD case, we demonstrate that 
the TS relaxation rates vary more than two orders of magnitude by
tuning the inter-dot barrier. This offers a feasible scheme to
 manipulate the TS relaxation in double QDs. The relaxation rates 
also sensitively depend on the dot size and magnetic field.

\begin{acknowledgments}
This work was supported by the Natural
Science Foundation of China under Grant Nos.\ 10574120 and 10725417,
 the
National Basic Research Program of China under Grant
No.\ 2006CB922005, the Knowledge Innovation Project of Chinese Academy
of Sciences and SRFDP.
\end{acknowledgments}

\begin{appendix}
  \section{$Q$ in Coulomb interaction}\label{appen}
Following Ref.\ \onlinecite{wu1}, we obtain $Q$ in Eq.\ (\ref{eq14}) as
  \begin{eqnarray}
    \nonumber
    &&Q(N_1,N_2,N_1^\prime,N_2^\prime)=\int^{\infty}_0dk_{\|}k_{\|}P_{N_1,N_1^\prime}(k_{\|})P_{N_2^\prime,N_2}(k_{\|})\\
    &&\hspace{1cm}\times\int^{\infty}_{-\infty}dk_z\frac{W_{N_1,N_1^\prime}(k_z)W_{N_2^\prime,N_2}^\ast(k_z)}{k^2}\ ,
  \end{eqnarray}
  where $P_{N,N^\prime}$ and $W_{N,N^\prime}$ come from the lateral and vertical parts 
of the matrix
  element $\langle n,l,n_z|\exp(i{\bf k\cdot r})|n^\prime,l^\prime,n_z^\prime\rangle$,
  respectively.  $P$ is given by\cite{wu1}
  \begin{eqnarray}
    \nonumber
    &&P_{N,N^\prime}(k_{\|})=\sqrt{\frac{n!n^\prime!}{(n+|l|)!(n^\prime+|l^\prime|)!}}\mbox{exp}\left(-\frac{k_{\|}^2}{4\alpha^2}\right)\\
    \nonumber
    &&\hspace{1cm}\times\sum_{i=0}^{n^\prime}\sum_{j=0}^{n}C_{n^\prime,|l^\prime|}^{i}C_{n,|l|}^{j}\bar{n}!L_{\bar{n}}^{|l-l^\prime|}\left(\frac{k_{\|}^2}{4\alpha^2}\right)\\
    &&\hspace{1cm}\times\left(\mbox{sgn}(l^\prime-l)\frac{k_{\|}}{2\alpha}\right)^{|l^\prime-l|}\ ,
  \end{eqnarray}
  with $C_{n,l}^{i}=\frac{(-1)^i}{i!}\binom{n+l}{n-i}$ and
  $\bar{n}=i+j+(|l|+|l^\prime|-|l^\prime-l|)/2$. sgn$(x)$ represents the sign
  function.  $W$ reads
  \begin{equation}
    W_{N,N^\prime}=\langle n_z|\mbox{exp}(ik_zz)|n_z^\prime\rangle\ .
  \end{equation}

\end{appendix}


\begin{thebibliography}{0}
\bibitem{loss1} D. Loss and D. P. DiVincenzo, Phys. Rev. A {\bf 57},
  120 (1998).
\bibitem{taru}R. Hanson, L. P. Kouwenhoven, J. R. Petta, S. Tarucha, and L. M. K. Vandersypen,
arXiv:cond-mat/0610433, and references therein.

\bibitem{spintronics} {\em Semiconductor Spintronics and Quantum
  Computation}, edited by D. D. Awschalom, D. Loss and N. Samarth
(Springer-Verlag, Berlin, 2002); I. \u{Z}uti\'{c}, J. Fabian, and
S. Das Sarma, Rev. Mod. Phys. {\bf 76}, 323 (2004), and references therein.

\bibitem{hanson} R. Hanson, B. Witkamp, L. M. K. Vandersypen,
L. H. Willens van Beveren, J. M. Elzerman, and L. P. Kouwenhoven,
Phys. Rev. Lett. {\bf 91}, 196802 (2003).

\bibitem{amasha} S. Amasha, K. Maclean, Iuliana Radu,
  D. M. Zumb\"{u}hl, M. A. Kastner, M. P. Hanson, and A. C. Gossard,
  arXiv:cond-mat/0607110.
%Phys. Rev. Lett. {\bf 98}, 036802 (2007).

%\bibitem{merkulov} I. A. Merkulov, Al. L. Efros, and M. Rosen, Phys.
%  Rev. B {\bf 65}, 205309 (2002).


\bibitem{koppen2} F. H. L. Koppens, C. Buizert, K. J. Tielrooij,
  I. T. Vink, K. C. Nowack, T. Meunier, L. P. Kouwenhoven, and
  L. M. K. Vandersypen, Nature {\bf 442}, 766 (2006).


\bibitem{petta} J. R. Petta, A. C. Johnson, J. M. Taylor, E. A. Laird,
  A. Yacoby, M. D. Lukin, C. M. Marcus, M. P. Hanson, and
  A. C. Gossard, Science {\bf 309}, 2180 (2005).

\bibitem{koppen} F. H. L. Koppens, J. A. Folk, J. M. Elzerman,
  R. Hanson, L. H. Willems van Beveren, I. T. Vink, H. P. Tranitz,
  W. Wegscheider, L. P. Kouwenhoven, and L. M. K. Vandersypen, Science
  {\bf 309}, 1346 (2005).
\bibitem{petta2} J. R. Petta, A. C. Johnson, A. Yacoby, C. M. Marcus,
  M. P. Hanson, and A. C. Gossard. Phys. Rev. B {\bf 72}, 161301
  (2005).
\bibitem{sasaki} S. Sasaki, T. Fujisawa, T. Hayashi, and Y. Hirayama,
Phys. Rev. Lett. {\bf 95}, 056803 (2005).
\bibitem{meunier} T. Meunier, I. T. Vink, L. H. Willems van Beveren,
  K-J. Tielrooij, R. Hanson, F. H. L. Koppens, H. P. Tranitz,
  W. Wegscheider, L. P. Kouwenhoven, and L. M. K. Vandersypen,
 Phy. Rev. Lett. {\bf 98},
  126601 (2007).
\bibitem{Woods}L. M. Woods, T. L. Reinecke, and Y. Lyanda-Geller,
  Phys. Rev. B {\bf 66}, 161318 (2002).
\bibitem{Sousa} R. de Sousa and S. Das Sarma, Phys. Rev. B {\bf
  68}, 155330 (2003).
\bibitem{wu1}J. L. Cheng, M. W. Wu, and C. L\"u, Phys. Rev. B {\bf
    69}, 115318 (2004).
\bibitem{dani1}D. V. Bulaev and D. Loss, Phys. Rev. B {\bf 71},
  205324 (2005).
\bibitem{Golovach}V. N. Golovach, A. Khaetskii, and D.
  Loss, Phys. Rev. Lett. {\bf 93}, 016601 (2004).
\bibitem{Destefani}C. F. Destefani and S. E. Ulloa, Phys. Rev. B {\bf
    72}, 115326 (2005).
\bibitem{Jose}P. San-Jose, G. Zarand, A. Shnirman, and G. Sch\"{o}n,
  Phys. Rev. Lett. {\bf 97}, 076803 (2006).
\bibitem{Falko2} V. I. Fal'ko, B. L. Altshuler, and O. Tsyplyatyev,
  Phys. Rev. Lett. {\bf 95}, 076603 (2005).
  \bibitem{wu2} Y. Y. Wang and M. W. Wu, Phys. Rev. B {\bf 74}, 165312
  (2006).
\bibitem{peter1}P. Stano and J. Fabian, Phys. Rev. Lett. {\bf 96},
  186602 (2006).
\bibitem{peter2}P. Stano and J. Fabian, Phys. Rev. B {\bf 74}, 045320 (2006).
\bibitem{Westfahl} H. Westfahl. Jr., A. O. Caldeira,
  G. Medeiros-Ribeiro, and M. Cerro, Phys. Rev. B {\bf 70}, 195320
  (2004).

\bibitem{climente} J. I. Climente, A. Bertoni, G. Goldoni, M. Rontani,
  and E. Molinari, Phys. Rev. B {\bf 75}, 081303 (2007).


\bibitem{dani2} V. N. Golovach, A. Khaetskii, and D. Loss,
  arXiv:cond-mat/0703427.

\bibitem{climente2} J. I. Climente, A. Bertoni, G. Goldoni, M. Rontani,
  and E. Molinari, Phys. Rev. B {\bf 76}, 085305 (2007).

\bibitem{marian} M. Florescu and P. Hawrylak, Phys. Rev. B {\bf 73},
  045304 (2006).

\bibitem{dress} G. Dresselhaus, Phys. Rev. {\bf 100}, 580 (1955).

\bibitem{rashba} E. I. Rashba, Fiz. Tverd. Tela (Leningrad) {\bf 2},
  1224 (1960).

\bibitem{paget} D. Paget, G. Lample, B. Sapoval, and V. I. Safarov,
Phys. Rev. B {\bf 15}, 5780 (1977).

\bibitem{pikus} G. E. Pikus and A. N. Titkov {\em Optical Orientation}
(Berlin, Springer, 1984).
\bibitem{erlingsson} Sigurdur I. Erlingsson, Yuli V. Nazarov, and
  Vladimir I. Fal'ko, Phys. Rev. B {\bf 64}, 195306 (2001).

\bibitem{fock} V. Fock, Z.Phys. {\bf 47}, 446 (1928).

\bibitem{darwin} C. G. Darwin, Math. Proc. Cambridge Philos. Soc. {\bf
    27}, 86 (1930).

\bibitem{austing} D. G. Austing, S. Sasaki, K. Muraki, K. Ono,
  S. Tarucha, M. Barranco, A, Emperador, M. Pi, and F. Garcias,
  Int. J. Quantum Chem. {\bf 91}, 498 (2003).

\bibitem{lau} W. H. Lau and M. E. Flatt\'{e}, Phys. Rev. B {\bf 72},
  161311(R) (2005).
\bibitem{yakonov} M. I. D'yakonov and V. I. Perel',
  Zh. \'{E}ksp. Teor. Fiz. {\bf 60}, 1954 (1971)[Sov. Phys. JETP {\bf
    38}, 1053(1971)].

\bibitem{vogl} P. Vogl, {\em Physics of Nonlinear Transport in
  Semiconductors}, edited by D. K. Ferry, J. R. Barker, and
  C. Jacoboni, Nato Advanced Study Institute Series B52 (Plenum Press,
  New York, 1980).

\bibitem{mahan} G. D. Mahan, {\em Polarons in Ionic Crystals and Polar
Semiconductors}, edited by J. T. Devreese (North-Holland, Amsterdan,
  1972).

\bibitem{lei} X. L. Lei, J. L. Birman, and C. S. Ting,
  J. Appl. Phys. {\bf 58}, 2270 (1985).

\bibitem{bs} {\it Semiconductors}, Landolt-B\"ornstein, New Series, Vol. 17a, edited by O.
Madelung (Springer-Verlag, Berlin, 1987).


\bibitem{richards} D. Richards, B. Jusserand, H. Peric, and
  B. Etienne, Phys. Rev. B {\bf 47}, 16028 (1993).


\bibitem{scat} U. Bockelmann and G. Bastard, Phys. Rev. B {\bf 42},
  8947 (1990).

%\bibitem{hatano} T. Hatano, M. Stopa, and S. Tarucha, Science {\bf
%    309}, 268 (2005).

%\bibitem{jordan} Jordan Kyriakidis, M. Pioro-Ladriere, M. Ciorga,
%  A. S. Sachrajda, and P. Hawrylak, Phys. Rev. B {\bf 66}, 035320 (2002).
\bibitem{efig2} Under the basis for Fig.\ 2, the
    TS crossing/anticrossing occurs near $B=1.6$\ T (not
    shown). Therefore, the selection rule already works at $B>1.3$\ T.

\end{thebibliography}
\end{document}